\documentclass[12pt]{article}

\usepackage{epsf}
\usepackage[dvips]{graphicx}

\begin{document}

\begin{center}
{\bfseries CHARM PRODUCTION IN NN AND $\gamma N$ COLLISIONS}

\vskip 5mm

Egle Tomasi-Gustafsson$^{ \dag}$
\vskip 5mm

{\small
{\it DAPNIA/SPhN, CEA/Saclay, 91191 Gif-sur-Yvette Cedex, France }
\\
$\dag$ {\it
E-mail: etomasi@cea.fr
}}
\end{center}

\vskip 5mm

\begin{center}
\begin{minipage}{150mm}
\centerline{\bfseries Abstract}
The processes of open and hidden charm production in $NN$ collisions and of open charm production in $\gamma N$ collisions are studied. The near-threshold cross section is predicted and polarization phenomena are calculated in frame of an effective Lagrangian approach.
\end{minipage}
\end{center}

\vskip 10mm

\section{Introduction}

New  experimental facilities, planned in the near future \cite{GSI}, will be able to make detailed measurements of open charm production  in $pN$, $NA$ or $AA$ collisions, in a wide energy region, starting from threshold. The upgraded machine at Jefferson Laboratory will open the possibility to produce charmed particles at threshold in photoproduction experiments \cite{Chudakov}. 

One can expect that the physics of close-to-threshold $\Lambda_cD$-production is similar to $\Lambda K$-production and use SU(4) symmetry to make predictions in the charm sector.

Few or no experimental data exist at threshold for open charm hadro and photo-production on the nucleon. More data exist for $J/\Psi$ production, sometimes indirectly derived from nuclear reactions.

The theoretical activity, for inclusive charm production, is mainly focused on QCD-inspired approaches, and it cannot be easily extrapolated to the threshold region. Moreover exclusive processes cannot be described in QCD approaches without additional assumptions.
For a reliable description of the overall charm dynamics in $AA$ collisions, it is necessary to have a good parametrization of the  elementary processes of open charm production in $\pi N$ and $NN$ collisions. The parametrization of the energy behavior of the total cross section for different processes of $D$ and $\overline{D}$-production  for $\sqrt{s}\ge$ 10 GeV, generally used in simulation codes, cannot be applied near threshold as it violates the general threshold behavior for two or three particle production.

In this contribution we present some of the results recently obtained for the processes $N+N\to \Lambda_c(\Sigma_c)+\overline{D}+N$ \cite{Re02}, $N+N\to N+N+J/\Psi$ \cite{Re02c} in the threshold region and for $\gamma+N\to \Lambda_c(\Sigma_c)+\overline{D}+N$ \cite{ETG04,Re02b}. We derive, in model independent way, the spin structure of the matrix element for these processes. For quantitative predictions, we calculate the cross section and the polarization phenomena in the framework of a model based on meson exchanges in an effective Lagrangian (ELA) approach. ELA gives a very convenient frame for this study, as all the observables can be calculated with the help of few parameters, such as masses, coupling consants, and magnetic moments, which have a definite physical meaning.

\section{Open charm production in NN collisions}

An important characteristic of the threshold $NN$-dynamics is the strong correlation between the spin and isospin structures of the threshold matrix element for the $N+N\to \Lambda_c(\Sigma_c)+\overline{D}+N$ reactions. Using the isotopic invariance of the strong interaction and taking into account that $I(\Sigma_c)=1$, $I(\overline{D})=1/2$, one can express the matrix element for the seven different processes $N+N\to \Lambda_c(\Sigma_c)+\overline{D}+N$, in terms of three (complex) isotopic amplitudes, $A_{I_1 I_2}$, corresponding to the total isospin $I_1$ for the initial nucleons and the total isospin $I_2$ for the produced $\overline{D}N$-system:
\begin{equation}
\begin{array}{rrrr}
{\cal M}(pp\to \Sigma_c^{++}D^- p)=
&A_{11}&+\sqrt{2}A_{10}, &  \\
{\cal M}(pp\to \Sigma_c^{++}\overline{D^0} n)=
&A_{11}&-\sqrt{2}A_{10},&  \nonumber \\
{\cal M}(pp\to \Sigma_c^+\overline{D^0} p)=
&-\sqrt{2}A_{11}, & & \label{eq:amp} \\
{\cal M}(np\to \Sigma_c^{++}D^- n )=
&~A_{11}& &+\sqrt{2}A_{01},\nonumber \\
{\cal M}(np\to \Sigma_c^0\overline{D^0} p)=
&-A_{11}&& +\sqrt{2}A_{01},\nonumber\\
{\cal M}(np\to \Sigma_c^+ D^- p)=&&A_{10} &-A_{01}, \nonumber\\
{\cal M}(np\to \Sigma_c^+\overline{D^0} n)=&&-A_{10} &-A_{01}, \nonumber
\end{array}
\end{equation} 
Generally, each isospin structure contains the contribution of all possible spins. However, the interferences $A_{11}\bigotimes A_{01}^*$ and $A_{10}\bigotimes A_{01}^*$, vanish in the threshold region (after summing over the polarizations of the colliding nucleons) and we can derive the following relations, between the differential cross sections, which hold for any model, at threshold:
$$ \displaystyle\frac{d\sigma}{d\omega}(np\to \Sigma_c^{++}D^- n )= \displaystyle\frac{d\sigma}{d\omega}(np\to \Sigma_c^0\overline{D^0} p),~ \displaystyle\frac{d\sigma}{d\omega}(np\to \Sigma_c^+ D^- p)=\displaystyle\frac{d\sigma}{d\omega}(np\to \Sigma_c^+\overline{D^0} n).$$ 

Following the ideology of meson production in $NN$-collisions, we calculate the cross section for $NN$ collisions, in terms of $D$-meson exchange in $t$-channel, neglecting, as it was done for strange particle production, the light mesons exchange. The cross section can be expressed in terms of the isotopic amplitudes of elastic $\overline{D}N$ scattering, $a_0$ (singlet) and $a_1$ (triplet). Using isotopic invariance, one finds for the ratio  ${\cal R}_D$ of the cross sections of $\Lambda_c\overline{D}$-production for $np$ over $pp$ collisions:
\begin{equation}
{\cal R}_D=\displaystyle\frac{\sigma(np\to\Lambda_c^+\overline{D^0} n)}
{\sigma(pp\to\Lambda_c^+\overline{D^0} p)}=
\displaystyle\frac{\sigma(np\to\Lambda_c^+{D^-} p)}
{\sigma(pp\to\Lambda_c^+\overline{D^0} p)}=
\displaystyle\frac{1}{4}
\left(1+\displaystyle\frac{a_0^2}{3a_1^2}\right )>\displaystyle\frac{1}{4},
\label{eq:ratio}
\end{equation}
where the index $D$ underlines that this prediction is valid only in framework of $D$-exchange. These large isotopic effects can, in principle, be modified by initial (ISI) and final (FSI) state interaction. However, as the reaction threshold is quite large, one can assume that ISI are similar for $np$ and $pp$. Moreover, assuming $|a_0/a_1|\simeq 0.1$ as for elastic $KN$ scattering, FSI would also be negligible.

Based on phase space arguments, one can estimate the relative 
cross section of the processes 
$p+p\to\Lambda_c^+ +\overline{D^0} +p$ ($D-$exchange) and $p+p\to\Lambda +K^+ +p$ ($K-$exchange). Charm production turns out to be lower by three order of magnitude than strange particle production. This is mainly due to the difference in the reaction threshold and in the propagators, as well as to the necessary phenomenological form factors. Taking the experimental data on the total cross section for $p+p\to \Lambda+ K^++p$ \cite{Mo02}, one can predict the following energy dependence for the cross section of open charm production, valid near threshold:
\begin{equation}
\sigma(pp\to\Lambda_c^+ \overline{D^0}p)\simeq 0.2(Q/0.1 ~\mbox{GeV})^2\mbox{ nb}.
\label{eq:eq15}
\end{equation}

\begin{figure}[h]
\mbox{\epsfysize=10cm\leavevmode \epsffile{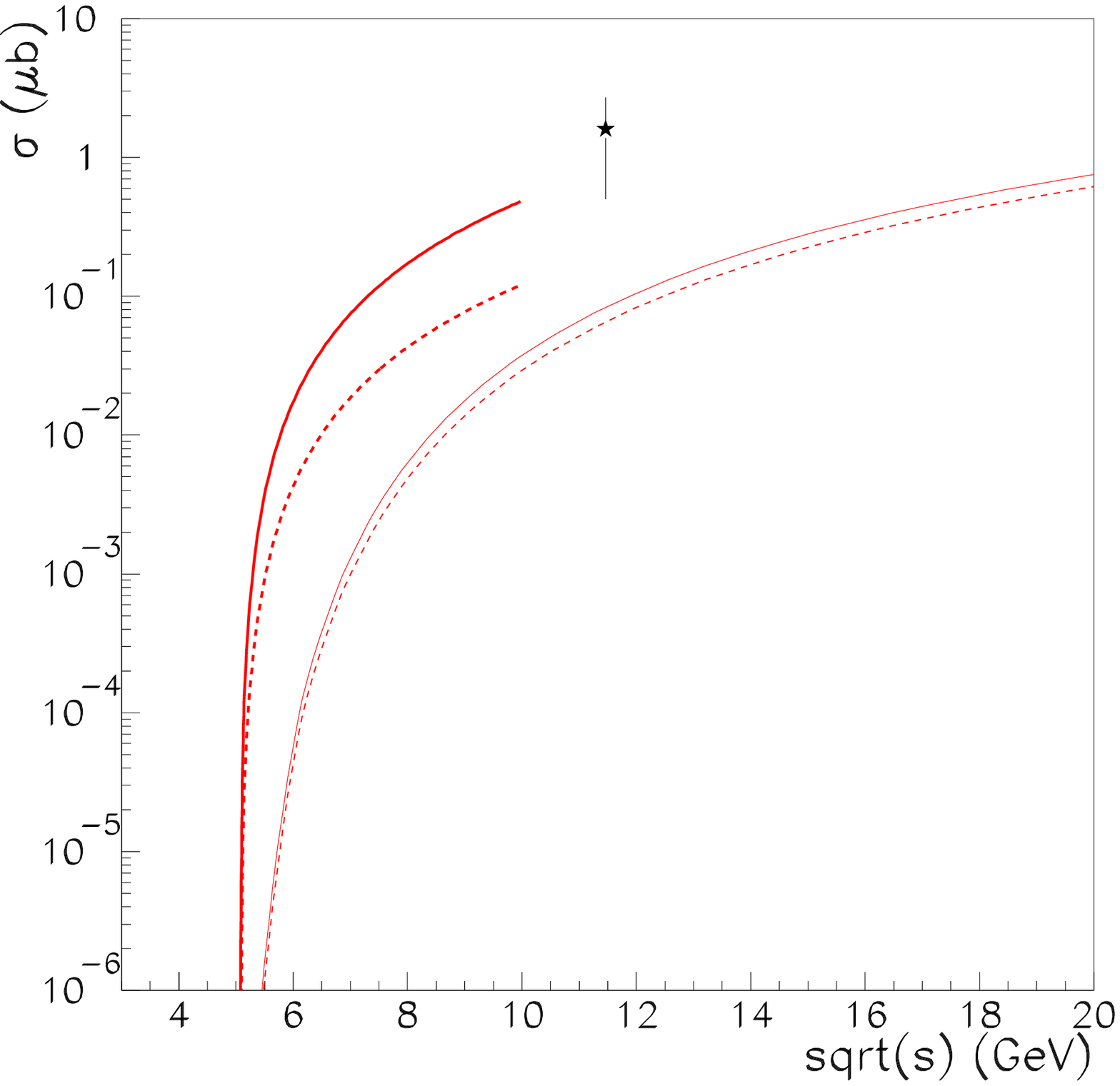}}
\label{fig:fig1}
\end{figure}

In Fig. \ref{fig:fig1} a comparison between the calculations \cite{ETG04} (thick lines)  and  \cite{Ca01} (thin lines) is shown for $\overline{D^0}$ production (solid lines) and for $D^-$-production (dashed lines). 

\vspace*{-1 true cm}
\vspace*{-1 true cm}
\begin{flushleft}
\begin{minipage}[t]{9 cm}
\noindent{\small 
{\bf Fig. 1} 
Comparison between the calculations \protect\cite{ETG04} (thick lines)  and  \protect\cite{Ca01} (thin lines) for $\overline{D^0}$ production (solid lines) and for $D^-$-production (dashed lines).
}
\end{minipage}
\end{flushleft}
\begin{flushright}
\vspace*{ -11 true cm}
\begin{minipage}[t]{6 cm}
One can see that the values can differ by an order of magnitude. This comes from the fact that the correct $Q$-behavior at threshold, from phase space considerations, should be quadratic, whereas the best fit parameters of Ref. \cite{Ca01} give an exponent $\simeq 5$. The difference is also due to the fact that the isotopic relations among $\overline{D^0}$ and $D^-$-cross sections are not respected in the parametrization \cite{Ca01}. This induces very large effect in particular in the threshold region. 
Note that the calculation \cite{Ga02}, which is based on the Quark-Gluon String model and the Regge phenomenology, is in agreement with \cite{ETG04}, for a specific choice of form factors. 
\end{minipage}
\end{flushright}

\section{$J/\Psi$ production in $NN$ collisions}

$J/\Psi$ production has a specific interest: the production and the propagation of charm in ion-ion collisions has been considered as one of the most promising probe of quark-gluon plasma (QGP) formation. The suppression of charmonium production in heavy ion collisions has been indicated as a signature of QGP \cite{Ma86}, but in order to state the evidence of a clear signal, it is necessary to analyze in detail all possible mechanisms for $J/\Psi$ production in ion-ion collisions, and also all other processes which are responsible for the dissociation of the produced $J/\Psi$ meson,  such as $J/\Psi+N\to \Lambda_c+D$, for example. The knowledge of the elementary process  $p+N\to p+N+J/\Psi$ is very important for a realistic calculation of $J/\Psi$ production in nucleus-nucleus collisions. 

In principle, the 'elastic' $J/\Psi$ production in $NN$-collisions can be treated in full analogy with processes of light vector meson production. All symmetry properties of the strong interaction, such as the Pauli principle, the isotopic invariance, the P-invariance, which have been successfully applied to light vector meson production in $NN$-collisions \cite{Re97}, hold for $J/\Psi$ production, too. A formalism can be built, which is particularly simplified in the threshold region, where all final particles are produced in $S$-state. Simple considerations indicate that this region may be quite wide: the effective proton size, which is responsible for charm creation, has to be quite small, $r_c\simeq 1/m_c$, where $m_c$ is the $c$-quark mass  \cite{Br01}. Therefore the $S$-wave picture can be applied for $q\le m_c$, where $q$ is the $J/\Psi$  three-momentum in the reaction center of mass (CMS). 
In the general case the spin structure of the matrix element for the process 
$N+N\to N+N+V$ is described by a set of 48 independent complex amplitudes, which are functions of five kinematical variables. The same reaction, in coplanar kinematics, is described by 24 amplitudes, functions of four variables. In collinear kinematics the number of independent amplitudes is reduced to 7 and the description of this reaction is further simplified in case of threshold $V$-meson production, where all final particles are in $S$-state.

Applying the selection rules following from the Pauli principle, the P-invariance and the conservation of the total angular momentum, it is possible to prove that the threshold process $p+p \rightarrow p+p+V^0$ is characterized by a single partial transition: 
\begin{equation}
S_i=1,~\ell_i=1~\to ~{\cal J}^{ P}=1^- \to S_f=0, 
\label{eq:eqpp}
\end{equation}
where $S_i$ ($S_{f}$) is the total spin of the two protons in the initial (final) 
states and $\ell_i$ is the orbital momentum of the colliding protons.
In the CMS of the considered reaction, the matrix element corresponding to transition (\ref{eq:eqpp}) can be written as:
\begin{equation}
{\cal M}(pp)=2f_{10}(\tilde{\chi}_2~\sigma_y ~\vec{\sigma}
\cdot\vec  U^* \times\hat{\vec k }\chi_1)~(\chi^{\dagger}_4 \sigma_y\ \tilde{\chi}^{\dagger}_3 )
\label{eq:mpp},
\end{equation}
where $\chi_1$ and $\chi_2$ ( $\chi_3$ and $\chi_4$) are the
two-component spinors of the initial (final) protons;  $\vec  U$ is the three-vector of the $V$-meson polarization, $\hat{\vec k}$ is
the unit vector along the 3-momentum of the initial proton; $f_{10}$ is the S-wave partial amplitude, describing the triplet-singlet transition of the two-proton system in V-meson production.

In case of  $np$-collisions, applying the conservation of isotopic invariance for the strong interaction, two threshold partial transitions are allowed:
\begin{equation}
S_i=1,~\ell_i=1~\to ~{\cal J}^{ P}=1^- \to S_f=0,~
S_i=0,~\ell_i=1~\to ~{\cal J}^{ P}=1^- \to S_f=1,
\label{eq:eqnp}
\end{equation}
with the following spin structure of the matrix element:
\begin{equation}
{\cal M}(np)=f_{10}(\tilde{\chi}_2~\sigma_y ~\vec{\sigma}
\cdot\vec  U^* \times\hat{\vec k }\chi_1)~(\chi^{\dagger}_4 \sigma_y\ \tilde{\chi}^{\dagger}_3 )+\nonumber \\
 f_{01}(\tilde{\chi}_2~\sigma_y\chi_1)(\chi^{\dagger}_4
\vec{\sigma}
\cdot\vec  U^* \times\hat{\vec k }\sigma_y \tilde{\chi}^{\dagger}_3 ).
\label{eq:mnp}
\end{equation}
Here $f_{01}$ is the S-wave partial amplitude describing the singlet-triplet transition of the two-nucleon system in V-meson production. In the general case the amplitudes $f_{10}$ and $f_{01}$ are complex functions, depending on the energies $E$, $E'$ and $E_V$, where $E,(E')$ and $E_V$ are the energies of the initial (final) proton and of the produced $V-$meson, respectively.

Note that $f_{10}$ is the common amplitude for $pp$- and $np$-collisions, due to the isotopic invariance of the strong interaction. This explains the presence of the coefficient two in Eq. (\ref{eq:mpp}). All dynamical information is contained in the partial amplitudes $f_{01}$ and $f_{10}$, which are different for the different vector particles. Some polarization phenomena show common characteristics, essentially independent from the type of vector meson. For example, vector mesons produced in $pp$- and $np$-threshold collisions are transversally polarized,
and the elements of the density matrix $\rho$ are independent from the relative values of the amplitudes $f_{01}$ and $f_{10}$: $\rho_{xx}=\rho_{yy}=\frac
{1}{2}$, $\rho_{zz}=0$.  

All other single spin polarization observables, related to the polarizations of the initial or final nucleons, identically vanish, for any process of $V-$meson production. The dependence of the differential cross section for threshold collisions of polarized nucleons (where the polarization of the final particles is not detected) can be parametrized as follows:
\begin{equation}
\displaystyle\frac{d\sigma}{d\omega}(\vec P_1,\vec P_2)=\left ( \displaystyle\frac{d\sigma}{d\omega}\right)_0
\left (1+{\cal A}_1 \vec P_1 \cdot \vec P_2 +{\cal A}_2 \hat{\vec k} \cdot\vec P_1\hat{\vec k} \cdot\vec P_2 \right ),
\label{eq:sig}
\end{equation}
where $\vec P_1 $ and $\vec P_2$ are the axial vectors of the beam and target nucleon polarizations, and $d\omega$ is the element of phase-space for the three-particle final state. The spin correlation  coefficients  ${\cal A}_1$ and ${\cal A}_2$ are real and they are different for $pp$- and $np$- collisions:
\begin{itemize}
\item $\vec p+\vec p\to p+p+V^0$:  ${\cal A}_1(pp)=0$, ${\cal A}_2(pp)=1$.
\item $\vec n+\vec p\to n+p+V^0$:  
${\cal A}_1(np)=-\displaystyle\frac{|f_{01}|^2}{|f_{01}|^2+|f_{10}|^2},~~{\cal
A}_2(np)=\displaystyle\frac{|f_{10}|^2}{|f_{01}|^2+|f_{10}|^2}$,
\end{itemize}
with the following relations $-{\cal A}_1(np)+{\cal A}_2(np)=1$ and 
$0\le {\cal A}_2(np)\le 1$.

Defining ${\cal R}$ as the ratio of the total (unpolarized) cross section for $np$- and $pp$- collisions, taking into account the identity of final particles in $p+p\to p+p+V^0$, we find:
\begin{equation}
{\cal R}=\displaystyle\frac{\sigma(np\to npV^0)}
{\sigma(pp\to ppV^0)}=\displaystyle\frac{1}{2}+\displaystyle\frac{1}{2}
\displaystyle\frac{|f_{01}|^2}{|f_{10}|^2},~\mbox{and }~
{\cal A}_1=-1+\displaystyle\frac{1}{2{\cal R}}.
\label{eq:eqr}
\end{equation}
Similar expressions for the  polarization transfer from the initial neutron to the final proton can be given in terms of the partial amplitudes $f_{01}$ and $f_{10}$.

One can show that the most probable mechanism to describe the threshold dynamics of $J/\Psi$-production in NN-collisions is meson exchange in $t$-channel,  with ${\cal J}^P=0^+$ (scalar mesons) and ${\cal J}^P=0^-$ (pseudoscalar meson).

For the exchange of one meson, $\pi$, $\rho$ or $\eta$, 
one finds that only one amplitude enters in the cross section, and the observables take definite values (Table I).

\begin{table}[h]
\begin{center}
\begin{tabular}{|c|c|c|c|}
\hline
&$\pi(f_{01}=-3f_{10})$&$\eta(f_{01}=f_{10})$& $\sigma(f_{01}=-f_{10})$\\ 
\hline\hline
$R $&5&1& 1\\ 
\hline
${\cal A}_1$ &-9/10&1/2& -1/2\\ 
\hline
${\cal P}_1$ &-3/5&-1& 1\\ 
\hline \hline
\end{tabular}
\caption{ Numerical values of the isotopic ratio and single spin polarization observables for $\pi$, $\eta$ or $\sigma$ exchange.}
\end{center}
\end{table}
\vspace*{-.5 true cm}
If one considers the exchange of two mesons, $\pi+\eta$, for example, one can still express the observables in terms of a parameter, $r$, which characterizes the relative role of $\pi$ and $\eta$ exchange and contains coupling constants, propagators and kinematical factors:
$$r=\displaystyle\frac{g_{\eta NN}}{g_{\pi NN}}\displaystyle\frac{h_{1\eta}}{h_{1\pi}}\left (\displaystyle\frac {t-m^2_{\pi}}{t-m^2_{\eta}}\right ).$$
The single spin polarization observables in $n+p\to n+p+V^0$ and  
the ratio ${\cal R}$ of the total cross section for $n+p$- and $p+p$-collisions can be written as a function of $r$ as:  
\begin{equation}
{\cal A}_1=-\displaystyle\frac{9+6{\cal R}e ~r +|r|^2}{2(5+2{\cal R}e ~r +|r|^2)}, 
~{\cal P}_1=-\displaystyle\frac{3-2{\cal R}e ~r -|r|^2}{5+2{\cal R}e ~r +|r|^2}, ~{\cal R}=\displaystyle\frac{5+2{\cal R}e ~r +|r|^2}{|1-r|^2}, 
\label{eq:pos1}
\end{equation}
where $h_{1\eta}$ ($h_{1\pi}$) is the partial amplitude which describes the threshold spin structure for the subprocess ${\pi^0}^*(\eta^*)+N\to V^0+N$ at threshold.

In order to make an estimation of the cross section, one can compare $\phi$ and $J/\Psi$ production, in framework of a model based on $\rho$ exchange. One finds that
$$\sigma(pp\to pp\phi)\simeq 206\left ( \displaystyle\frac{Q}{0.1 \mbox{~GeV~}}\right ) ^2\mbox{~nb~}$$ 
and 
\begin{equation}
\sigma(pp\to pp J/\Psi)\simeq  9.7\cdot 10^{-5}\left ( \displaystyle\frac{Q}{0.1 \mbox{~GeV~}}\right ) ^2 \left [F(t_{J/\Psi})/F(t_{\phi})\right ]^2 \mbox{~nb~}.
\label{jpsi}
\end{equation}
This value is too small, when compared with the existing experimental value for the lowest $\sqrt{s}=6.7$ GeV, namely $\sigma_{exp}(pp\to ppJ/\Psi)=0.3\pm 0.09$ nb. To explain this discrepancy, one can note that the $\rho$-exchange model for $\sigma(\pi N\to J/\Psi)$ gives a cross section one order of magnitude smaller in comparison with other possible theoretical approaches \cite{Bo75,Ko79,Be76}.  Another possibility is to take
$\left [F(t_{J/\Psi})/F(t_{\phi})\right ]^2\simeq 10$, which can be plausible, because the $J/\Psi=c\overline{c}$-system must have a smaller size in comparison with $\phi=s\overline{s}$. This can be realized by the following form factor:
$$F_V(t)= \displaystyle\frac{1}{1-{t}/{\Lambda_V^2}},~{\Lambda_V}\simeq m_V.$$

The cross section, based on Eq. (\ref{jpsi}) normalized the the experimental point at $\sqrt{s}=6.7$ GeV, i.e., taking the ratio 
$\left [F(t_{J/\Psi})/F(t_{\phi})\right ]^2=10$, is plotted in Fig. 2, together with the experimental data from the compilation \cite{Vogt}, where different symbols differentiate $J/\Psi$ production in $pp$ or extrapolated from $pA$ collisions. Note, that, for $r$ real and $\simeq 0$,
\vskip -1.2 true cm
\begin{figure}[h]
\mbox{\epsfysize=10.cm\leavevmode \epsffile{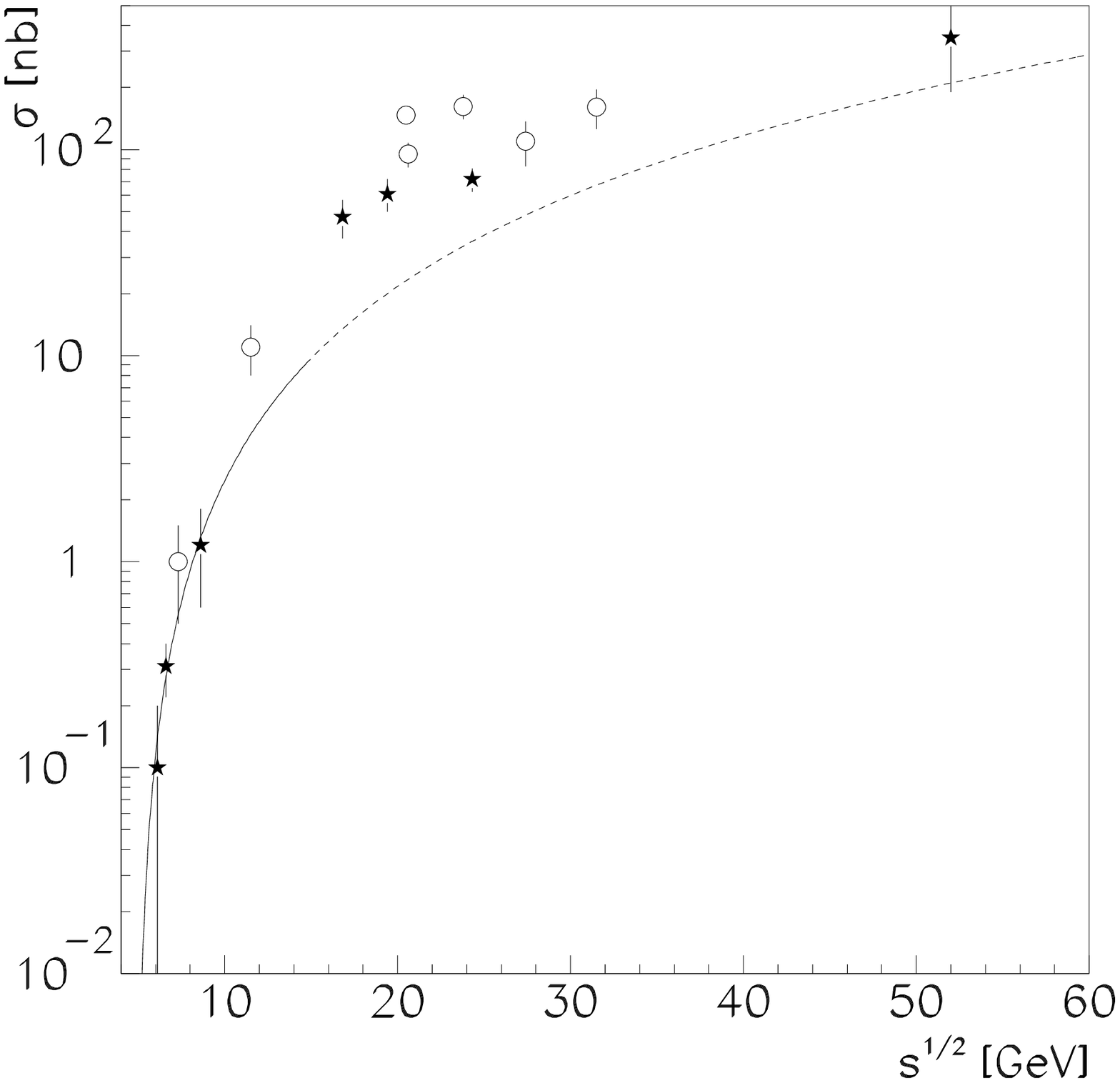}}
\end{figure}
\vspace*{-0.8 truecm}
\begin{flushleft}
\begin{minipage}[b]{9 cm}
\noindent{\small{\bf Fig. 2} Cross section for $J/\psi$ production in $pp$ collisions. Data are from \protect\cite{Vogt}. 
}
\end{minipage}
\end{flushleft}
\begin{flushright}
\vskip -10.6 true cm
\begin{minipage}[b]{6.1 cm}

\noindent one finds for the isotopic ratio, ${\cal R}=5$,
which would require a correction of the experimental data on $pA$ reactions, where equal $np$ and $pp$ cross sections are usually assumed for the extraction of the elementary cross section.
The reactions of $J/\Psi$ production in $pp$ and $np$ collisions present very different characteristics concerning: the number of independent partial transitions, the spin structure of the threshold matrix elements, the value of the absolute cross sections and the polarization phenomena.
All these differences are generated by a common mechanism: the different role of the Pauli principle for $pp$ and $pn$ collisions in the near the threshold region. 
\end{minipage}
\end{flushright}
\vspace*{-0.3 truecm}
The experimental determination of the ratio of the total cross sections for $np$ and $pp$collisions is very important for the identification of the reaction mechanism.

\section{$\gamma$-nucleon collisions}
We give here model independent considerations for the reaction $\gamma+p\to \Lambda_c^+ 
+\overline{D}^0$, where the spins involved are $1+1/2\to 1/2+ 0$. In collinear kinematics, due to the conservation of the spin projection, the collision of $\gamma$ and $p$ with parallel spins cannot take place for collinear regime. Therefore, the asymmetry in the collision of circularly polarized photons with a polarized target, takes its maximum value. This result holds for any process of pseudoscalar and scalar meson photoproduction on a nucleon target (if the final baryon has spin 1/2) and it is independent on the ${\cal P}$-parity of produced meson. It is a model independent result, based uniquely on the assumption of the spins of the particle, the conservation of helicity in collinear kinematics, and applies also to the threshold region, for any angle.

One can suggest a model independent method to determine the parity of the charmed mesons, more exactly, the relative  $P(N\Lambda_cD)$ parity. 
We will use the standard parametrization \cite{Ch57} of the spin structure for the amplitude of pseudoscalar meson photoproduction on the nucleon:
\begin{equation}{\cal M}(\gamma N\to Y_c\overline{D}_c)=\chi_2^{\dagger}\left [i\vec\sigma\cdot\vec e f_1 + 
\vec\sigma\cdot\hat{\vec q}\vec\sigma\cdot\hat{\vec k}\times\vec e f_2+ 
i\vec e \cdot\hat{\vec q}\vec\sigma\cdot\hat{\vec k}f_3
+i\vec\sigma\cdot\hat{\vec q} \vec e \cdot\hat{\vec q}f_4\right ]\chi_1,
\label{eq:mat}
\end{equation}
where $\hat{\vec k}$ and
$\hat{\vec q}$ are the unit vectors along the three-momentum of $\gamma$ and 
$\overline{D}_c$; $f_i$, $i$=1,4 are the scalar amplitudes, which are functions of two independent kinematical variables, the square of the total energy $s$ and $\cos\vartheta$, where $\vartheta$ is the $\overline{D}_c$--meson production angle in the reaction center of mass (CMS) with respect to the direction of the incident photon,  $\chi_1$ and  $\chi_2$ are the two-component spinors of the initial nucleon and the produced $Y_c$-baryon. Note that the pseudoscalar nature of the $\overline{D}_c$--meson is not experimentally confirmed up to now, therefore Eq. (\ref{eq:mat}) corresponds to the prescription of the quark model for the {\cal P}-parities of $\overline{D}_c$ and $Y_c$-charm particles. 

It is possible to derive in a model independent way, that for the reaction $\gamma+p\to \Lambda_c^+ +\overline{D}^0$, the triple polarization correlation coefficient is sensitive to $P(N\Lambda_cD)$. In collinear kinematics, the spin structure (\ref{eq:mat}) reduces to one amplitude, which depends on $P(N\Lambda_cD)$:
\begin{eqnarray}
{\cal F}^{(-)}_{col}&=&\vec\sigma\cdot\vec e f^{(-)}_{col},\nonumber \\
{\cal F}^{(+)}_{col}&=&\vec\sigma\cdot\vec e\times\hat{\vec k} f^{(+)}_{col}, 
\label{eq:mppf} 
\end{eqnarray}
where $f^{(\pm)}_{col}$ is the collinear amplitude for $P(N\Lambda_cD)=\pm 1$.
Due to the presence of a single allowed amplitude in collinear kinematics, all polarization observables have definite numerical values, which are independent on the model chosen for $f^{(\pm)}_{col}$.

The dependence of the $\Lambda_c$ polarization on the polarization of the colliding particles can be written as:
$$-(\vec e\cdot\vec e)(\vec P_1\cdot\vec P_2)+2(\vec e\cdot\vec P_1)(\vec e\cdot\vec P_2), \mbox{ ~if~ }  P(N\Lambda_cD)=-1,$$
\begin{equation}
(\vec e\cdot\vec e)[(\vec P_1\cdot\vec P_2)-2(\hat{\vec k}\cdot\vec P_1)(\hat{\vec k}\cdot\vec P_2)]-
2(\vec e\cdot\vec P_1)(\vec e\cdot\vec P_2), \mbox{~if~} P(N\Lambda_cD)=+ 1,
\label{eq:par} 
\end{equation}
where $\vec P_1$ and $\vec P_2$ are the polarization vectors for the initial and final baryons.

One can see from (\ref{eq:par}), that only the linear photon polarization affects the triple polarization correlations in $\vec\gamma+\vec p\to \vec\Lambda_c^++\overline{D^0}$, due to the P-invariance of the electromagnetic interaction of charmed particles. Let us define the coordinate system for the considered collinear kinematics with  the $z-$axis  along $\hat{\vec k}$ and the $x-$ axis  along the vector ${\vec e}$ of the photon linear polarization. The correlations  (\ref{eq:par}) can be written in such system as:
$$-(\vec P_1\cdot\vec P_2)+2 P_{1x} P_{2x}= P_{1x} P_{2x}-P_{1y} P_{2y} -P_{1z} P_{2z}, \mbox{ ~if~ }  P(N\Lambda_cD)=- 1,$$
\begin{equation}
(\vec P_1\cdot\vec P_2)-2 P_{1z} P_{2z}-2P_{1x} P_{2x}=-P_{1x} P_{2x}+P_{1y} P_{2y} -P_{1z} P_{2z}, \mbox{~if~} P(N\Lambda_cD)=+ 1.
\label{eq:mpp2} 
\end{equation}

From (\ref{eq:mpp2}) one can find a connection between the components of the vectors $\vec P_1$ and $\vec P_2$ for the different $P(N\Lambda_cD)$, assuming,  for simplicity, that initially one has 100\%  linearly polarized photons:
$$P_{2x}= +P_{1x},~ P_{2y}=-P_{1y}, P_{2z}=- P_{1z}, \mbox{ ~if~ } 
P(N\Lambda_cD)=- 1,$$
\begin{equation}
P_{2x}= -P_{1x},~ P_{2y}=+P_{1y}, P_{2z}=- P_{1z}, \mbox{~if~} P(N\Lambda_cD)=+ 1,
\label{eq:mpp3} 
\end{equation}
One can see that both transversal components of the $\Lambda_c$-polarization are sensitive to $P(N\Lambda_cD)$, through the relative sign between $P_{2i}$ and $P_{1i}$:
\begin{equation}
P_{2x}=-P(N\Lambda_cD)P_{1x},~P_{2y}=P(N\Lambda_cD)P_{1y},
\label{eq:sol} 
\end{equation}
whereas $P_{2z}=- P_{1z}$ for any value of $P(N\Lambda_cD)$.

Therefore, the relations ($\ref{eq:sol}$) allow one to determine, in a  model-independent way, the $D-$meson P-parity. This model independent result holds for any nucleon photoproduction of a spin 1/2 baryon and a spin zero meson. It requires the assumption on P-parity conservation in $\gamma+p\to \Lambda_c^++\overline{D^0}$ and helicity conservation in the collinear regime. Similar arguments apply to the reaction $\vec\gamma+\vec p\to\Theta^++K$ and allow to suggest a method for the determination of the parity of pentaquark \cite{Re04}. 

The suggested experiment, measuring the triple polarization correlations in 
$\vec \gamma+\vec p\to \vec \Lambda_c^++\overline{D^0}$, can be in principle realized by the Compass collaboration \cite{COMPASS}, which  has a polarized target and where linearly polarized photons can be obtained in muon-proton collisions, at small photon virtuality, by tagging the photon through the detection of the scattered  muon.
The $\Lambda_c$-polarization can be measured through the numerous weak decays of the  $\Lambda_c^+$-hyperon, for example $\Lambda_c^+\to \Lambda +e^+ +\nu_e$ \cite{PdG}, which is characterized by a large decay asymmetry. In other words, the $\Lambda_c$ is a self-analyzing particle. 

Eqs. (\ref{eq:sol}) show that only the relative sign of the transversal components of the polarization of the target proton and the produced  
$\Lambda_c^+$-hyperon is important for the determination of the 
$P(N\Lambda_cD)$-parity. Therefore such experiment does not need very large statistics, only well identified events. The energy of the photon beam has not to be necessarily monochromatic. 

The charm particle photo and electroproduction at high energy is usually interpreted in terms of photon-gluon fusion, $\gamma+G\to c+\overline{c}$. Near threshold, other possible mechanisms, based on the subprocess $q+\overline{q}\to G\to c+\overline{c}$ should also be taken into account. In case of exclusive reactions, such mechanism is equivalent to the exchange of a $\overline{c}q$-system, in $t$-channel \cite{Od87}. The mesonic equivalent of such exchange is the exchange by pseudoscalar $\overline{D}_c$ and (or) vector $\overline{D}_c^*$ mesons, in the $t$-channel of the considered reaction.

In order to insure the gauge invariance, baryonic exchanges in $s$- and $u$-channels have also to be taken into account. Due to the virtuality of the exchanged hadrons, in this approach, form factors (FFs) are introduced in the pole diagrams.

\begin{figure}[h]
\mbox{\epsfysize=10.cm\leavevmode \epsffile{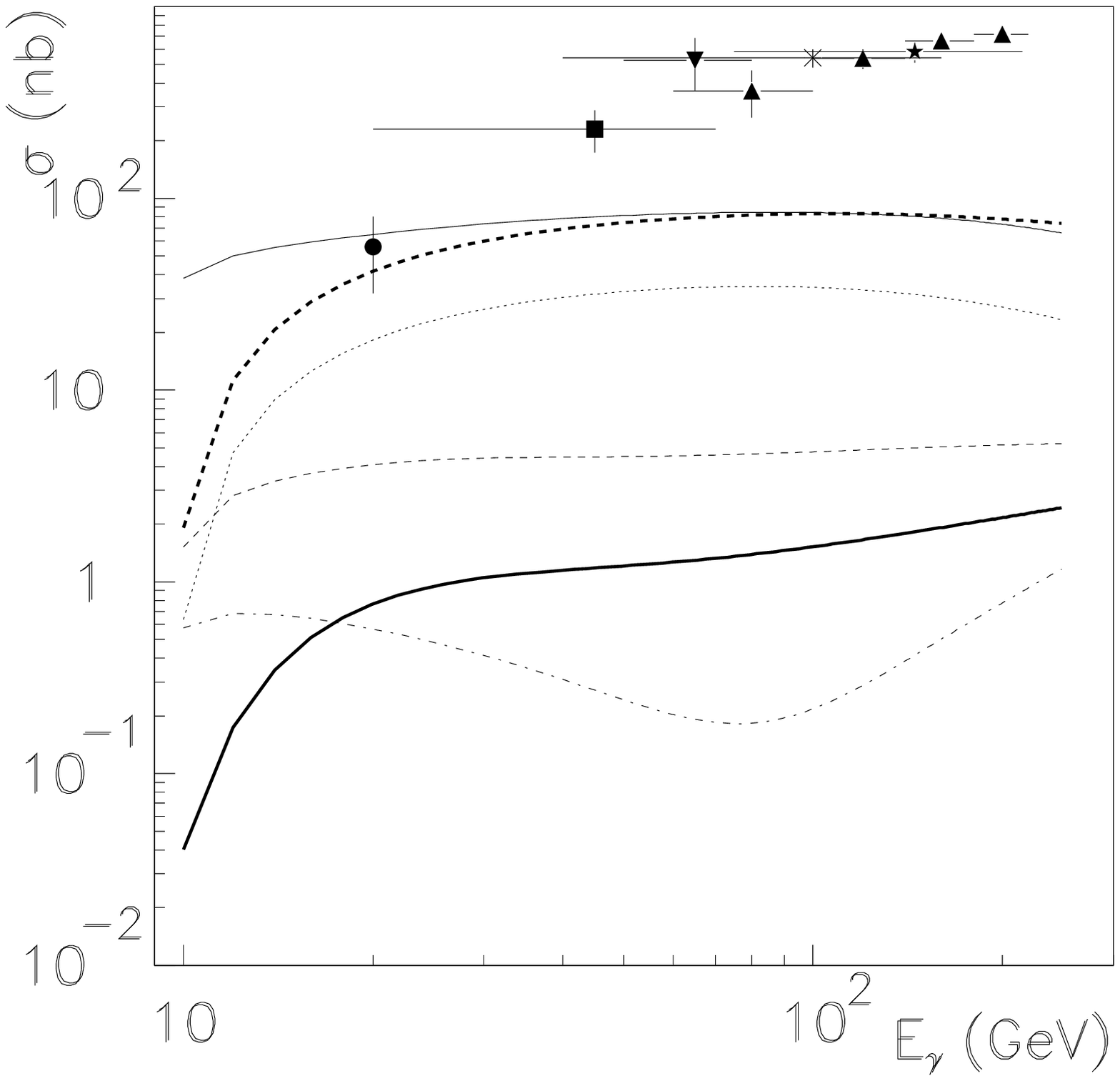}}
\label{fig:exc0}
\end{figure}

The  largest cross section on the neutron target belongs to the process  $\gamma +n\to \Sigma_c^0 +\overline{D}^0$, the $D^-$ production being essentially suppressed. The $D^-$ production is also small in the $\gamma p$ interaction,  $\gamma +p\to \Sigma_c^{++} +{D^-}$, in agreement with the experiment \cite{Ad87}. Large isotopic effects are an expected property of ELA approach, because the relative values of $s$, $u$, and $t$-channel contributions are different for the different channels (Fig. \ref{fig:exc0}). 

Polarization effects are generally large (in absolute value), characterized by a strong $\cos\vartheta$-dependence, which results from a coherent effect of all the considered pole contributions.

 Large isotopic effects (i.e. the dependence on the electric charges of the participating hadrons) are especially visible in the $\cos\vartheta$-distributions for all these observables. 
\vspace*{-1 true cm}
\vspace*{-9 true cm}
\begin{flushright}
\begin{minipage}[b]{6.5 cm}
{\small {\bf Fig. 3} $E_\gamma$-dependence of the total cross section for photoproduction of charmed particles for model I. The curves correspond to different reactions: $\gamma+p\to \Lambda_c^+ 
+\overline{D}^0$ (solid line),
$\gamma+p\to\Sigma_c^{++}+D^-$ (dashed line), 
$\gamma+p\to \Sigma_c^+  + \overline{D}^0$ (dotted line), 
$\gamma+n \to \Lambda_c^+ +D^-$  (dot-dashed line), 
$\gamma+n \to \Sigma_c^+  + D^-$ (thick solid line), 
$\gamma+n \to \Sigma_c^0  + \overline{D}^0$ (thick dashed line). The data correspond to the total charm photoproduction cross section ( \protect\cite{ETG04} and refs. herein)
}
\end{minipage}
\end{flushright}
\vspace*{1 true cm}
 In frame of the considered model, the asymmetry $\Sigma_B$ is positive in the whole angular region, in contradiction with the predictions of PGF \cite{Du80} and in agreement with the SLAC data \cite{Ab86}.

\section{Instead of Conclusions}

{\it This talk is dedicated to Prof. Michail P. Rekalo. The results presented here would not have been realized without his deep and wide knowledge in different fields of physics. I met him in for the first time in this Hall of the Bogoliubov Laboratory of Theoretical Physics, ten years ago. Since then, I could profit of many enthusiastic discussions and of his clear explanations.

Extremely creative, full of ideas, he was always ready to interact with people.  Very generous in scientific discussions, he had a prompt humour, sometimes very sharp.  

He was able to think deeply and to concentrate on physics. He could focus very quickly on the essential aspects of a problem: complicated and difficult questions were suddenly made very simple and solvable.

His absence is an unvaluable loss for our community.}

\end{document}